\documentclass[conference]{IEEEtran}
\IEEEoverridecommandlockouts
\usepackage{cite}
\usepackage{amsmath,amssymb,amsfonts}
\usepackage{algorithmic}
\usepackage{graphicx}
\usepackage{textcomp}
\usepackage{xcolor}
\def\BibTeX{{\rm B\kern-.05em{\sc i\kern-.025em b}\kern-.08em
    T\kern-.1667em\lower.7ex\hbox{E}\kern-.125emX}}
    
\usepackage{graphicx}
\usepackage{url}
\usepackage{enumitem}
\setlist[itemize]{leftmargin=*}
\usepackage{listings}
\usepackage{pgfplotstable}
\usepackage[ruled,norelsize]{algorithm2e}
\usepackage{hyperref}
\usepackage{lipsum}

\newenvironment{algoitemize}
{ \begin{itemize}
    \setlength{\itemsep}{0pt}
    \setlength{\parskip}{0pt}
    \setlength{\parsep}{0pt}     }
{ \end{itemize}                  }

\usepackage{tabularx,colortbl}
\usepackage{tikz}
\usetikzlibrary{patterns}

\begin{document}

\title{To test, or not to test:  A proactive approach for deciding complete performance test initiation}

\author{\IEEEauthorblockN{1\textsuperscript{st} Omar Javed}
\IEEEauthorblockA{\textit{Department of Informatics, Universit\`a della svizzera italiana} \\
Lugano, Switzerland \\
omar.javed@usi.ch}
\and
\IEEEauthorblockN{2\textsuperscript{nd} Prashant Singh}
\IEEEauthorblockA{\textit{Department of Information Technology, Uppsala University} \\
Uppsala, Sweden \\
prashant.singh@scilifelab.uu.se}
\and
\IEEEauthorblockN{3\textsuperscript{rd} Giles Reger}
\IEEEauthorblockA{\textit{School of Computer Science, University of Manchester} \\
Manchester, United Kingdom \\
giles.reger@manchester.ac.uk}
\and
\IEEEauthorblockN{4\textsuperscript{th} Salman Toor}
\IEEEauthorblockA{\textit{Department of Information Technology, Uppsala University} \\
Uppsala, Sweden \\
salman.toor@it.uu.se}

}

\maketitle

\begin{abstract}
Software performance testing requires a set of inputs that exercise different sections of the code to identify performance issues. %Identifying performance issues is challenging because they are not easily observable. 
However, running tests on a large set of inputs can be a very time-consuming process. It is even more problematic when test inputs are constantly growing, which is the case with a large-scale scientific organization such as CERN where the process of performing scientific experiment generates plethora of data that is analyzed by physicists leading to new scientific discoveries. %that generates a plethora of data used by physicists to conduct experiments leading to new scientific discoveries. 
Therefore, in this article, we present a test input minimization approach based on a clustering technique to handle the issue of testing on growing data. Furthermore, we use clustering information to propose an approach that recommends the tester to decide when to run the complete test suite for performance testing. To demonstrate the efficacy of our approach, we applied it to two different code updates of a web service which is used at CERN and we found that the recommendation for performance test initiation made by our approach for an update with bottleneck is valid.
\end{abstract}


\section{introduction}
performance issues during the early stage of software development is crucial to avoid software failures~\cite{performancebugs}. It has been shown that performance bugs can cause critical software failures that can result in the abandonment of big projects~\cite{1901Cens91}. Furthermore, performance bugs are often very costly to diagnose because their consequences are not easily observable, which leads to many hours of diagnosis and fixing~\cite{performancebugs}.  

One approach to performance bug detection is unit testing. However, anecdotal evidence shows that unit testing for performance analysis is not nearly as common as unit testing of code correctness (i.e., functional bug detection)~
\cite{perftesteval}. One of the reasons is that performance issues arise only with particular inputs. %conditions. 
Hence, they can easily escape into production. 

To overcome this issue, studies have shown that profiling unit tests with different inputs can lead to identifying performance bugs~\cite{perflearner, performancebugs}. Furthermore, these studies use inputs that are gathered from analyzing bug patches. However, they do not consider large-scale input sets used by real users.
%These studies use inputs that are synthetically generated (i.e., inputs gathered from bug patches).  However, they do not consider large-scale input sets used by real users. 

 A big challenge for performance testing is to identify interesting input space i.e., inputs that can expose performance bugs~\cite{perflearner}. This makes performance testing over a large set of inputs a very time-consuming process.    %This is because developers execute tests on all different inputs to evaluate the performance of the software on code changes (i.e., performance regression testing). 
The reason is that unit tests are executed repeatedly on different inputs to evaluate the performance of the software on code changes (i.e., performance regression testing). 

Furthermore, the high testing time can be more problematic for systems in organizations that deals with generation of growing data. For example, CERN (the European Organization for Nuclear Research) processes prodigious amounts of data every day (in the order of Petascale)~\cite{storage}. Moreover, the Large Hadron Collider (LHC) which is the ``most powerful particle accelerator'' facilitate different detectors such as the Compact Muon Solenoid (CMS), ATLAS etc in carrying out experiments,   %which is one of the experiments at CERN 
produces a large volume of data i.e., 90 PetaByte each year~\cite{storage}. 
The issue for such system is the dependency on the stored data that grows over time. Therefore, it is necessary for performance testing to account for that growth.

To avoid the overhead and turnaround time involved in
checking for performance issues on code change. Existing studies~\cite{mathias,metric} have proposed different strategies for optimizing performance regression testing based on either test selection or code metrics. However, these studies do not consider the effect of data growth. %effect of growing inputs.

Therefore, in this article, we use a web service, employed by the CMS experiment at CERN called Conditions Uploader service~\cite{cmsuploader} as a use case
%we use a web services at LHC called Conditions Uploader service~\cite{cmsuploader}  as a use case  %The web service deals with correction and validation of scientific data. This makes it very significant because the correct and efficient functionality allows scientists to access and conduct experiments on the data. 
to demonstrate the problem of performance regression testing with increasing number of test inputs. We use the Conditions uploader service because it adds a level of complexity to the traditional performance testing paradigm i.e., the data growth problem. Here, the growth problem is  because of the continued running  of  LHC  that further  generates  more  data. This growing data  is used as test inputs for the testing of CMS uploader service.   
 Consequently,  making  performance  regression  testing on increasing inputs a challenging task.

\subsection{Motivating example}

\begin{figure}[h!]
	\centering
	\includegraphics[scale=0.35]{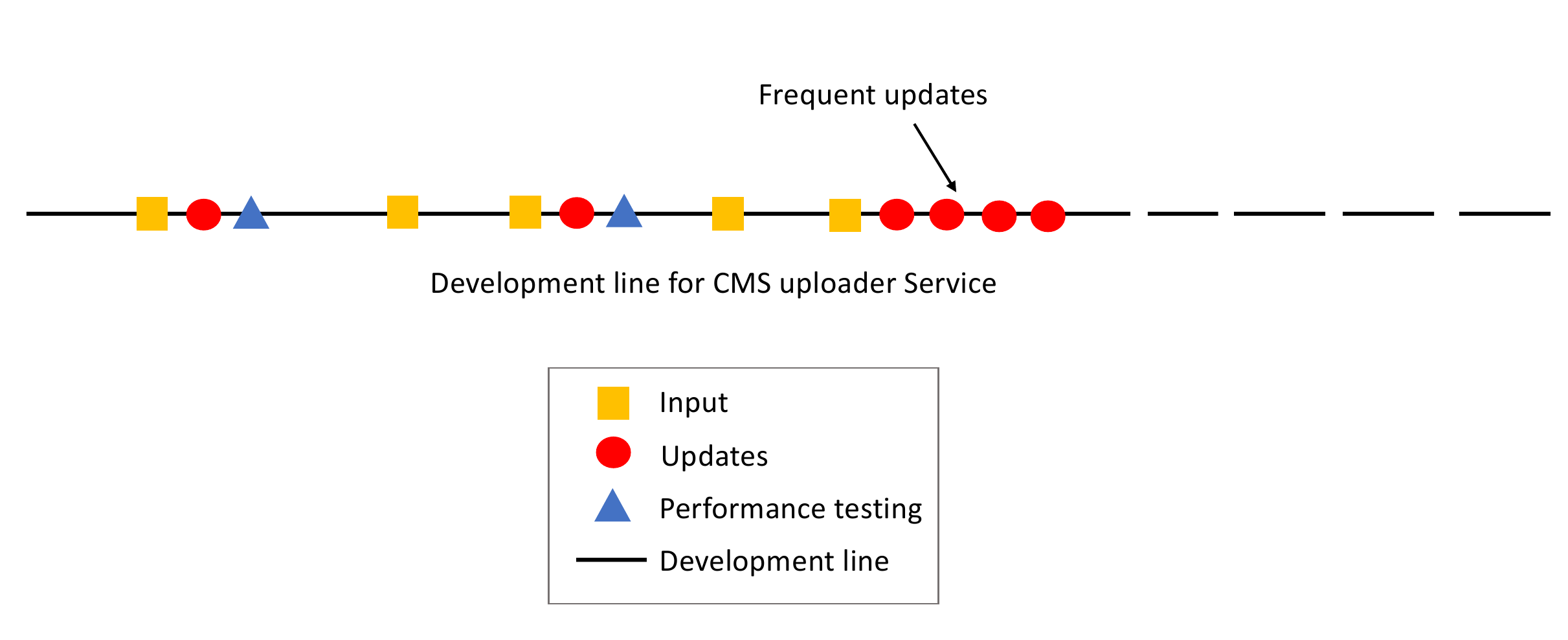}
	\caption{Figure showing the development timeline of CMS uploader service. }
	\label{fig:timeline}
\end{figure}

\begin{figure}[h!]
\centering
	\includegraphics[scale=0.35]{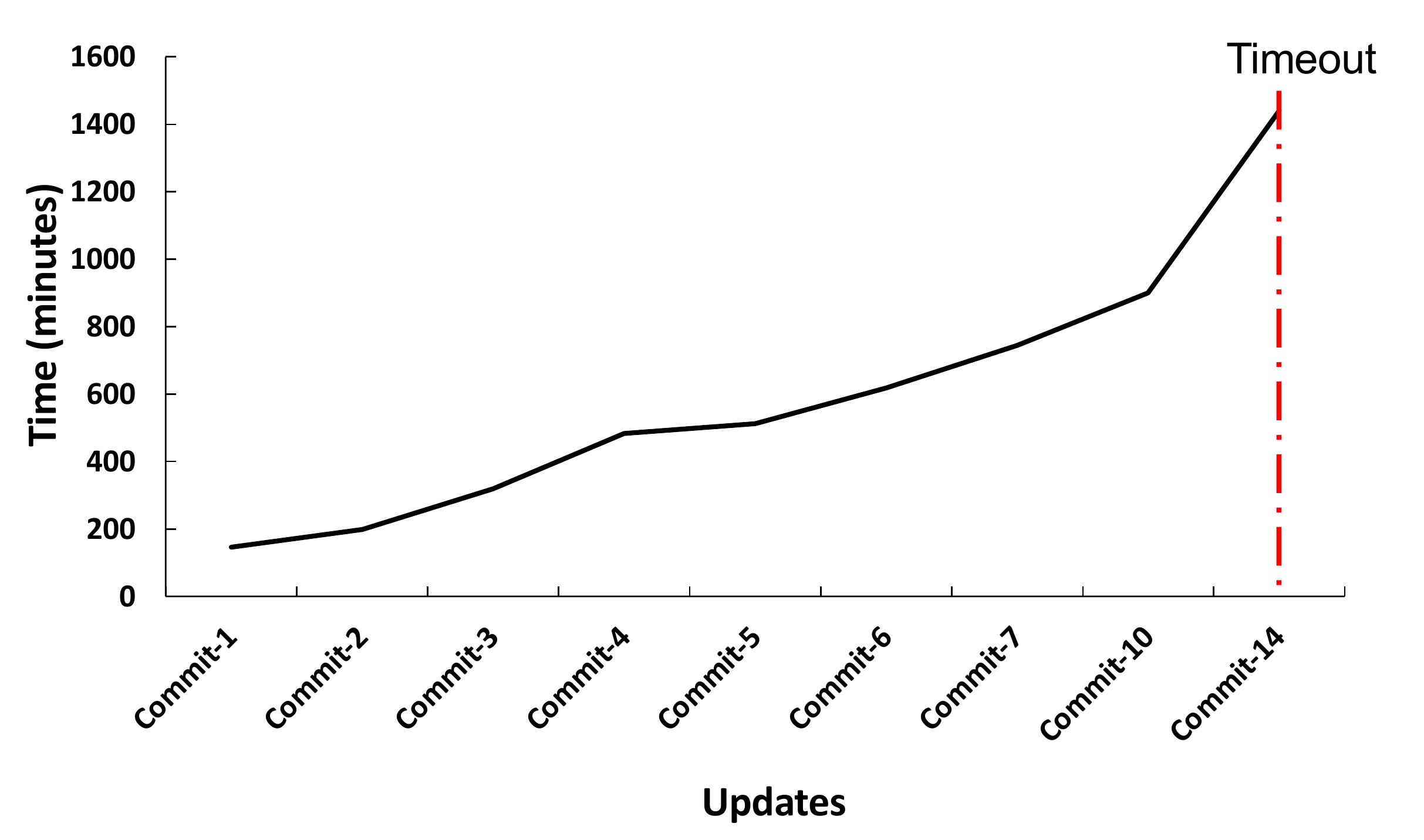}
	\caption{Figure demonstrating performance testing with increasing inputs. It shows the increase in execution time of testing with the increase in inputs at each commit. Commit-1 is the oldest update whereas Commit-14 is the latest. }
	\label{fig:timeout}
\end{figure}

Figure~\ref{fig:timeline} and \ref{fig:timeout} uses the CMS uploader service to demonstrate two key problems of performance regression testing with growing data. The problem is demonstrated in the context of Continuous Integration (CI) which is now a widely used practice, as it performs automated builds and regression tests (i.e., functional and non-functional) of software.  We discuss these two problems as follows:

\subsubsection{Frequent Update Problem}
The first problem is the execution of a test suite on frequent updates (as demonstrated by red circles in the Figure~\ref{fig:timeline}). The figure provides an illustration of the practice of running a test suite on frequent updates to check for performance bugs. This practice is often impractical and challenging for developers, since the waiting time for performance tests is high. Therefore, developers run tests in batches i.e., testing is done on a weekly or monthly basis or when there is an important update~\cite{perfweektest}. This infrequent execution of the tests makes it difficult for developers to detect which update caused the issue~\cite{perfweektest}.

\subsubsection{Data growth Problem}
The second problem is the execution of the test suite after the addition of new data (or test inputs) leading to a higher execution time; this can be seen in Figure~\ref{fig:timeout} which represents the actual part of update history of CMS uploader web service. The figure highlights the time taken when increasing data i.e., number of inputs at each update. %To handle this problem, testers can select a subset of random data on each regression testing. However, this will further complicate testing as developers will need to keep track of which inputs are used during testing in order to avoid selecting similar inputs on each regression testing.
Furthermore, we observe from the figure that regression testing with growing test inputs can quickly lead to a timeout in CI; this is because existing  CI services limit the use of resources for a longer duration of time.  

Based on this premise, we propose an approach to overcome the aforementioned problems of performance regression testing.

\subsection{Contribution of the article}

We present three key contributions of this study, which are as follows:

\begin{itemize}

%\item We present an evaluation of container vulnerability scanning tools by analyzing 59 popular public Docker images. 

\item[1.] We discuss and address the issue of (ever) growing data in the context of web service testing. This study is timely since many organizations are facing (or will face) this imminent issue.  
%We propose a minimization approach (based on unsupervised learning).

\item[2.] We demonstrate test input minimization approach. %based on clustering to identify similar inputs. 
With our approach, testers can overcome the problem of high test execution time with growing data. 

\item[3.] We propose a recommendation approach to help developers in deciding when it is feasible to initiate performance testing. This recommendation approach can address the problem of costly test execution at each update.

%We propose an approach to recommend testers to initiate complete performance testing. Our approach uses information from clusters to make a decision about performance regression testing. This can address the problem of costly test execution at each update.

%\item We provide a CI analytics system that improves the existing Continuous integration process with decision making. This will not be helpful for testers of CMS uploader service at CERN but also for developers for Python-based projects.

%We demonstrate the feasibility of our approach on a web service which is used at one of the experiments Large Hadron Collider in largest particle physics laboratory CERN.

%\item We identified and demonstrate a set of features which are useful for minimizing inputs and test case. Our features are not limited to CERN CMS webservice. These can be used for future studies related to code level predictive analysis.

\end{itemize}

%In summary, we aim to  (a) optimize performance testing on a very large input data set by considering two aspects: 1) identifying suitable unit tests for the analysis 2) input space reduction, and (b) improve existing mechanism of CI process to allow developers to easily understand and detect performance issues in the early phase of software development.

\subsection{Structure of the article}
%To help readers understand the flow of our article, we divide it into separate sections, which are as follows:

The article is divided into separate sections, which are as follows:

\begin{itemize}
% \item Section 2 introduces reader with the case study of CERN CMS uploader and its background in terms of performance testing.

% \item Section 3  provides necessary concepts and definitions which are used throughout the article.

\item Section~\ref{sec:background} presents the background and related work in this area of study.

\item Section~\ref{sec:approach} explains our approach for the recommendation of performance regression testing.

\item Section~\ref{sec:evaluation} and ~\ref{sec:recommendation} presents evaluation and our findings.

\item Section~\ref{sec:conclusion} provides our conclusion for the study.

\item In appendix, we present our analytic system for CI and also significance test.

\end{itemize}
\subsection{Resources}

We provide implementation of our algorithms for recommending performance testing, which are available in a public repository website~\cite{slowdown}. Furthermore, we also provide source code of our CI analytic system as well as the instructions for setting up the front-end and back-end of our analytic system. This can also be found in a public repository~\cite{cianalytic}.

\section{Background and related work}
\label{sec:background}
In this section, we provide a background of the CMS uploader service. Furthermore, we explore the existing state-of-the-art in the area of performance regression tests  and  input selection. Finally we show existing CI tools and services as well as highlight CI tools used in performance regression testing.

\subsection{CERN CMS Uploader service}

Data growth is becoming a big problem for systems in organizations that deals with ever growing data. One such example is CERN which has world's largest and most powerful particle accelerator known as Large Hadron Collider (LHC). Furthermore, the Compact Muon Solenoid (CMS) is a general-purpose detector at one of the interaction points on the LHC~\cite{cms}. Physics analysis requires a process of reconstruction that %consists 
makes use of collision events as well as alignment and calibration or ``conditions'' data.  During LHC runs, conditions data begins with its computation and ends with its upload to a central Conditions database.

The service responsible for uploading conditions data is the CMS Uploader service. It performs correctness checks before conditions data are uploaded. The CMS uploader service has 65 test cases and approximately 40K test inputs representing conditions data that continues to grow due to LHC runs. Therefore, adequate test size, growing input space, and its usage by real users at CERN make the CMS uploader service a reasonable use case for our study. 

Furthermore, the importance of automated performance analysis and testing over different inputs for the CMS uploader service has already been demonstrated in previous works ~\cite{joshuavypr,perfci}. In this study, we will address the issues and challenges of performance regression testing of the CMS uploader service with growing data. 

\subsection{Regression performance test selection}

Traditional regression test selection techniques focus mostly on correctness tests. There are few studies that focus on performance regression test selection~\cite{mathias, perfjit, metric}. These studies use previous commit information on making test selection. In contrast, our approach helps developers in deciding when to run performance tests. % recommendation on the initiation of performance testing,  
Our approach is based on approximating the slowdown of the system by sampling minimal information of code behavior with respect to test inputs. Therefore, our approach, does not require historical code change information to make predictions, rather uses information about current code modifications to make a decision.

Furthermore, there are studies based on heuristics to identify potential performance changes e.g., assigning costs to source code changes~\cite{sourcecodelearn}, prioritizing performance tests based on the impact a code change has on the system's performance~\cite{perfranker} and selection of test inputs to determine a performance model (e.g. the function of the response time)~\cite{iris,selectionperf}. We draw inspiration from these studies to build a recommendation approach that identifies regression performance test opportunities.

\subsection{Input selection for testing}

%PerfLearner~\cite{perflearner} generates different inputs by identifying important parameters from the bug report. 

Existing study on input selection~\cite{perflearner} tries to understand the effectiveness of performance bug detection. A specific set of inputs is %are 
generated based on the information present in the bug report. To select inputs based on program behavior, Qi Luo~\cite{inputprof} proposed an approach that analyzes the performance behavior of the program and tries to connect the behavior with test input data. This work is similar to ours in terms of selecting inputs based on program behavior. %However, we minimize inputs that show similar behavior of the program based on the test execution.  
%Furthermore, %there are different test input prioritization approaches  proposed on a learning system such as Deep Neutral Networks (DNN). 
Furthermore, one of the state-of-the-art approach DeepGini~\cite{deepgini}  prioritizes test inputs by measuring the confidence of  Deep Neutral Networks (DNN)  for classifying each test input. Test inputs are prioritized higher based on similar probabilities for all classes. Moreover, a study~\cite{boosttest} aims to estimate the accuracy of the DNN model by selecting a small set of test inputs. Compared to these studies, we use unsupervised learning to reduce the number of test inputs that exercise similar program behavior based on unit test execution.
%allow developers to select subset of inputs. We provide the flexibility of developers to select different learning algorithm based on how well the model behaves for their specific project.

\subsection{Continuous Integration services}
Continuous integration (CI) is the practice of integrating different code changes by multiple developers. It consists of automatically checking correctness of the code before integration. There are a number of popular CI options available such as CircleCI~\cite{circleci}, TravisCI~\cite{travisCI}, Codeship~\cite{codeship}, Bitbucket pipelines~\cite{bitbucket}, SemaphoreCI~\cite{semaphore}, and many more. These CI tools allow developers to automatically execute unit tests (in a remote server) before pushing changes to the code on a  code repository service like GitHub. Hence, a CI process provides a separation of concern (i.e., one service handles code version control, while a different service handles fault detection). 

However, many existing CI services only provide functionality for correctness tests. There are separate tools and plugins that support performance testing in CI e.g., Jenkins uses the Taurus tool for load testing~\cite{jenkins}. PeASS~\cite{peass} is a tool which identifies performance degradation between two code versions. Similarly, PerfCI~\cite{perfci} is a toolchain for automated performance testing which allows developers to specify the number of test inputs to be used for performance testing. Our approach could be helpful in such scenarios where the cost of performance tests is quite high due to a large number of inputs. 

\section{Approach for recommending performance tests}
\label{sec:approach}

\begin{figure}[ht!]
    \centering
	\includegraphics[scale=0.32]{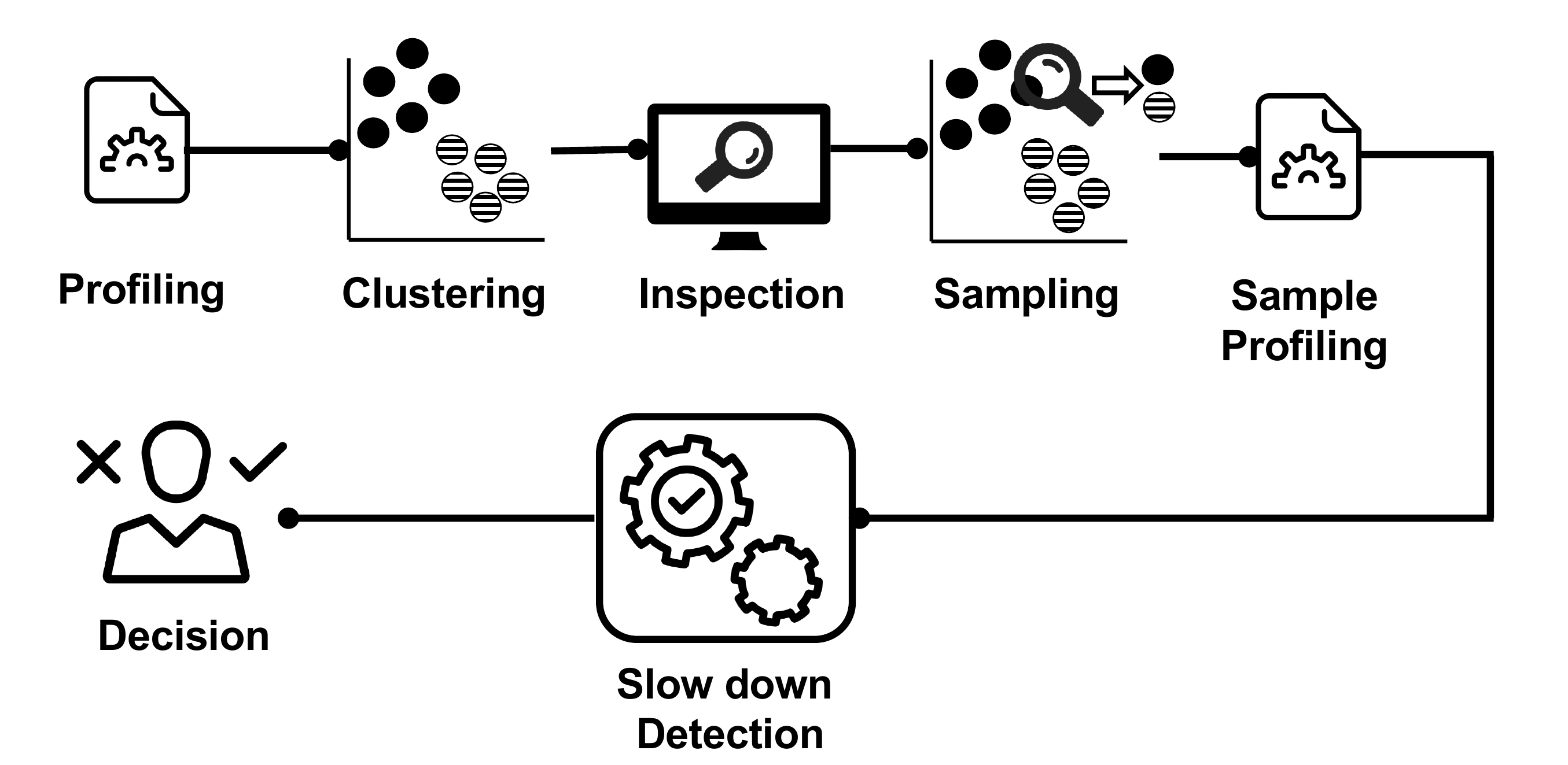}
	\caption{Our approach for recommending performance test initiation }
	\label{fig:diagram}
\end{figure}

This section describes our approach for recommending performance test initiation. Our approach consists of several stages as shown in figure~\ref{fig:diagram}. Therefore, we %first begin by explaining 
explain different stages that work together to assist developers in making informed decisions about test initiation. %Furthermore, we explain how our decision algorithm functions which is the core part of our recommendation approach.

Our approach works on two different types of updates; 1) addition of test inputs and 2) code modifications. Profiling, clustering and inspection are three stages that are applied to cluster test inputs. On the other hand, sampling, sample profiling and slow down detection  are applied on code updates.

%Initially, our approach minimizes test inputs and stabilizes the clustering information. The information of data points in clusters are used by our decision algorithm evaluate whether there is a need for initiating performance tests or not.

\begin{figure*}[h!]
	\centering

	\includegraphics[scale=0.6]{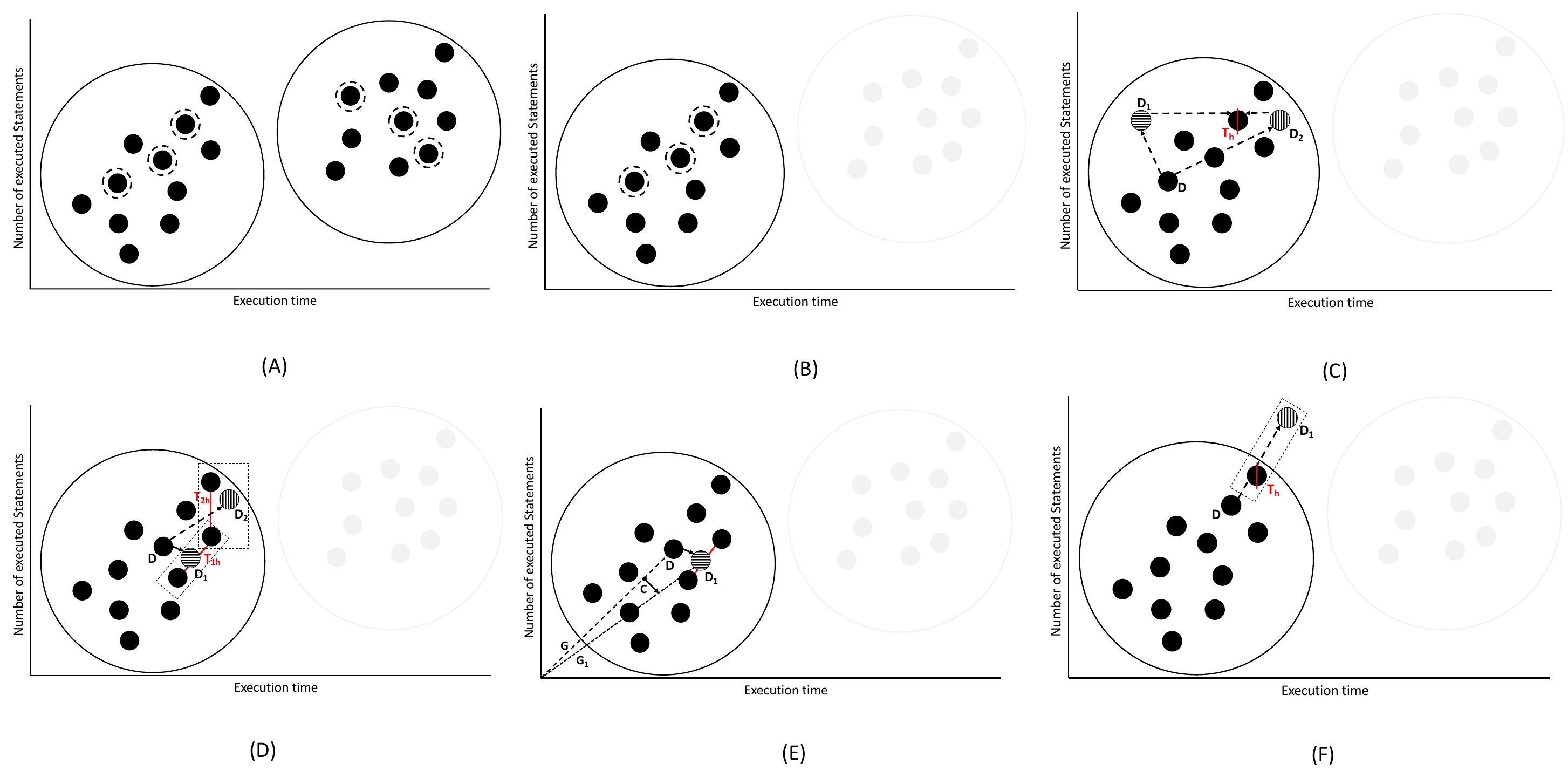}
	\caption{Figure demonstrating different steps taken by our approach for recommending performance regression testing. Fig.3 A) shows sampling of datapoints in clusters, B) selects cluster (in normal black color) one at a time, C) \& D) identification of time threshold, E) gradient measurement and F) handling of outlier datapoint.}
	\label{fig:algo}
\end{figure*}

\subsection{Profiling}

%We apply static and dynamic analysis to observe program behavior and collect program information such as execution time, memory usage, size of the input etc., representing different attributes. These attributes correspond to an observation (or a data point). Furthermore, our single observation corresponds to an application code executed by a test suite (i.e., code behavior) on a single test input. A test suite consists of several test cases; each test case executes a part of the application code. 

We apply static and dynamic analysis to observe program behavior and collect different program information such as execution time, memory usage, size of the input etc. In machine learning terminology, we refer to each program information (such as execution time, memory usage etc) as attributes. 

A test suite consists of several test cases; each test case executes a part of the application code. Therefore, the test suite is executed repeatedly on different test inputs to exercise different section of the application code.

Therefore, profiling (or code analysis) is performed during test execution to apply unsupervised machine learning algorithms to group test inputs based on similar attributes. %group code behavior as a function of test inputs based on similar attributes. 
The selection of attributes (or program information) is based on our experience with performance analysis of CMS uploader service and from previous test case minimization studies~\cite{mintest}. We further explain each one of the attributes as follows:

%This is done to identify different attributes that will be used by the learning model. The features are identified based on experiment and from previous test case minimization studies˜\cite{}. We perform clustering at two different levels of test data i.e., one is done for input and the other one for test cases. These two levels also share similar features which are explained as follows:

\begin{itemize}
    \item \textit{Execution time:} The total time it takes for the test suite to complete execution on each test input. 
    
    \item \textit{Memory usage:} The memory used during the execution of the test suite on each test input. 
    
    \item \textit{Total number of iterations:} %This is the total number of loop iterations that are executed on each input. 
    This represents the accumulated number of loop iterations executed in different sections of the code exercised by the test cases in a test suite on each test input.
    
    \item \textit{Total statements executed:} Statements which are executed by the test suite on each test input.
    
    \item \textit{Function calls:} Function calls made during the execution of the test suite on each test input.
    
    \item \textit{Conditional executed:} Sequence of conditions taken in the execution of the test suite on each test input. %The conditional paths taken in the execution of a test suite on each test input. %Condition if, else if or else conditions.
    
    \item \textit{Input Size:} Input represent various aspects of the CMS detector's configuration. Therefore, the size of each input varies based on the information it contains about the detector configuration.
    %Conditions data has different information about the detector. Therefore, each condition data contains different size; this condition data is passed as test inputs to the test suite.
    
    %\item \textit{Test result:} The result of each test case. The test can pass or fail based on whether the code fulfills the test requirement.
\end{itemize}

\subsubsection{Minimizing measurement perturbations}

One of the issues of profiling program's resources such as execution time is measurement perturbation. Instrumentation inserted in the code can perturbate measurements resulting in incorrect measurement of execution time~\cite{perturbation}.
%This is due to to the overhead of the instrumentation which is inserted in the code~\cite{perturbation}. 
%Hence, perturbation results in incorrect execution time measurement. 
To overcome this issue, we use PerfCI~\cite{perfci} to carry out our profiling task. PerfCI allows developers to  write pluggable  user-defined  analysis  for  CI. It  leverage stages~\cite{stages} in a CI process to allow different  profiling analysis  to  execute  separately. Therefore, one can measure resources such as execution time in a different stage and other attributes can be collected in a separate stage. Furthermore, stages can be executed in parallel to quickly complete profiling and ensure that profiling itself does not lead to perturbations of profiling results.

\subsection{Clustering}

Identifying similar code behavior by clustering has been carried out in previous software testing studies~\cite{testcluster, kmeancluster, fuzzycluster}. However, our study addresses the issue of performance regression testing in growing test inputs. Furthermore, clustering of test inputs provides the benefit of gaining useful insights about the test inputs with respect to code behavior. This information would be useful for developers when selecting test inputs.   

%However, the problem this study is tackling is different than the ones addressed before; we are trying to optimize performance testing under the issue of (ever) growing test inputs. Furthermore, it provides the benefit of gaining useful insights about the test inputs with respect to code behavior. This information would be useful for developers when selecting test inputs.   

Reduction of test inputs is carried out by clustering data points collected from profiling. These data points represents program behavior on each test input. After clustering of test inputs, there is a stage called inspection which allows developer to check whether the clustering applied is appropriate.

%To ensure stable clustering, we only update test inputs i.e., increase the number of test inputs. We avoid code modifications since it will change data points resulting in change of number of clusters, thereby affecting cluster stability which in turn, will affect our decision algorithm.

\subsection{Inspection}

It is not always easy to identify clear clusters. The order in which data points are arranged may affect the clusters. Furthermore, another issue with identification of clear clusters is with  data which has many missing values~\cite{clusterissues}. 

The arrangement of initial clusters is also important.  A careful and comprehensive  analysis  of  data  is  required.  If  the  initial clusters are not carefully and properly chosen, then after some iterations, clusters may become empty~\cite{clusterissues}. Therefore, cluster inspection is necessary to ensure correct clustering is done. In this regard,  different clustering algorithms and parameters are used for correct clustering. In appendix, we present our dashboard that allows inspection of clustering data.

We now explain sampling, sample profiling and slow down detection with the help of fig~\ref{fig:algo} along with algorithm ~\ref{algo:decision} and ~\ref{algo:eval}. This makes the core decision making part of our approach.

\begin{algorithm}[htb]
\label{algo:decision}
  \caption{Decision to Perform Testing}
   Decide(\textit{$D_u$}:updated data point, \textit{$T_h$}: time threshold,\\ \hspace{7ex} \textit{C}: current cluster, \textit{$A_l$}: acceptable limit) \\

  \begin{algoitemize}
    \item[1.]
    Initialize decision $D$ to False
    \item[2.]
    \textbf{IF} time information \textit{$T_u$} in \textit{$D_u$} $>$ \textit{$T_h$}
    \item[3.] \hspace{1ex} Set $D$ to True
    \item[4.]
    \textbf{ENDIF}
    \item[5.]
        \textbf{IF} $D$ is not True
    \item[6.]
        \hspace{1ex} Get time information $T_p$ in previous data point \\ \hspace{1ex} from current cluster \textit{C}
    \item[7.]
        \hspace{1ex} Get no. of executed statement information $S_p$ in \\ \hspace{1ex} previous data point from current cluster \textit{C}
      \item[8.] \hspace{1ex} Get time information  \textit{$T_u$} in  \textit{$D_u$}
      \item[9.]
        \hspace{1ex} Get no. of executed statement information \textit{$S_u$} in \textit{$D_u$}
      \item[10.]
        \hspace{1ex} Calculate gradient \textit{$G$} on points (0, $T_p$) , (0,$S_p$)
      \item[11.]
        \hspace{1ex} Calculate gradient \textit{$G_1$} on points (0, $T_u$) , (0,$S_u$)
      \item[12.]
        \hspace{1ex} Calculate gradient change \textit{$G_c$} between \textit{$G$} and \textit{$G_1$}
      \item[13.]
    \hspace{1ex} \textbf{IF}  \textit{$G_c$} $>$ \textit{$A_l$}
     \item[14.] \hspace{2ex} Set $D$ to True
     \item[15.]
    \hspace{1ex} \textbf{ENDIF}
    \item[16.]
    \textbf{ENDIF}
    \item[17.]
    Return $D$
  \end{algoitemize}
\end{algorithm}

\subsection{Sampling}
\label{sec:sample}
After clustering of inputs, our approach is prepared for identifying performance regression testing opportunities based on code updates. We use the number of executed statements and execution time as the two key attributes for decision making. Intuitively, these two attributes should exhibit correlation as executed statements has an impact on the execution time~\cite{srctime}. We further demonstrate the correlation of these attributes in the evaluation section.

Figure~\ref{fig:algo}A shows selection of data points by random sampling.  Sampling is done to collection minimum program information based on new code updates. Therefore, we kept the sampling size at a minimum e.g., we use 3 data points from each cluster. These data points contain information about specific test inputs which are used by sample profiling.

%minimum information about the code based on code updates  which is based on new updates. Therefore, sampling size is kept at a minimum e.g., we use 3 data points from each cluster. 

\subsection{Sample profiling}

In this stage, profiling is done with sampled test inputs on code update. Program information such as "number of executed statements" and "execution time" is collected in this stage.  Further optimization can be achieve in sample profiling. Instead of running all tests on sampled test inputs, one can execute test cases which exercise application code that has been updated. This can be achieved by identifying difference between commits. A \textbf{git diff} command~\cite{gitdiff}  can be used for this purpose.

%After data points are sampled, tests can be executed with the new updates. Profiling is done to collect the new state of the program but with only sampled inputs that is selected from sampling. This is a quick check for decision making. 

\begin{algorithm}[htb]
\label{algo:eval}
  \caption{Cluster data point evaluation}
  DataPointEvaluation(\textit{$C_l$}: list of clusters,  \textit{$D_d$}: list of \\ \hspace{20ex}  updated data points)
  \begin{algoitemize}
    \item[1.]
    \textbf{FOR EACH } \textit{c} in \textit{$C_l$}
    \item[2.]
    \hspace{1ex}\textbf{FOR EACH } \textit{$d$} in \textit{$D_d$}
    \item[3.]
    \hspace{2ex} \textbf{IF} num. of executed statements \textit{$s_e$} in \textit{$d$} equal to \textit{$s_e$} \\ \hspace{2ex} in \textit{c}
    \item[4.] \hspace{4ex} Get time information $t$ for $s_e$ in \textit{$c$}.
    \item[5.] \hspace{4ex} return Decide (\textit{$d$}, $t$, \textit{$c$})
    \item[6.]
    \hspace{2ex}    \textbf{ELSE} 
    \item[7.]
        \hspace{4ex} Get data points above and below the datapoint \textit{$d$} \\ \hspace{4ex} in  cluster \textit{$c$}
    \item[8.]
        \hspace{4ex} \textbf{IF} datapoint above (or below) \textit{$d$} does not exist
      \item[9.] \hspace{6.5ex} Get max time information $t_m$ for datapoint \\ \hspace{6.5ex} below  (or above) in cluster \textit{c} 
      \item[10.] \hspace{6.5ex} return Decide (\textit{$d$}, $t_m$, \textit{$c$})
      \item[11.] \hspace{4ex} \textbf{ELSE}
    \item[12.] \hspace{6.5ex} Find max time information $t_a$ of datapoint in \textit{$c$}
     \\ \hspace{6.7ex} above \textit{$d$}
     \item[13.] \hspace{6.5ex} Find max time information $t_b$ of datapoint in \textit{$c$} 
     \\ \hspace{6.7ex} below \textit{$d$}
     \item[14.] \hspace{6.7ex} Calculate mid-point $t_{md}$ of $t_a$ and $t_b$ 
     \item[15.] \hspace{6.7ex} return Decide (\textit{$d$}, $t_{md}$, \textit{$c$}) 
     \item[16.] \hspace{4ex} \textbf{ENDIF}
     \item[16.] \hspace{2ex} \textbf{ENDIF}
     \item[17.] \hspace{1ex} \textbf{END FOR}
     \item[18.]  \textbf{END FOR}
     
  \end{algoitemize}
\end{algorithm}

\subsection{Slow down detection and decision making}

After sampling and sample profiling stage, our algorithm maps new updated data points to their respective clusters. This mapping is done by selecting one cluster at a time as shown in figure~\ref{fig:algo}B, where selected cluster is highlighted in black. 

To detect a slow-down, our algorithm initially tries to find a time threshold value for the updated data point. Therefore, it attempts to detect the location of the updated data point in its cluster. We demonstrate this in figure~\ref{fig:algo}C, if the "number of executed statements" attribute of the updated data point matches with the "number of executed statements" attribute in any previous data point. Our algorithm tries to find a data point that has the highest execution time" attribute  among all data points sharing the same "number of executed statement" attribute. This will be adjudged as the time threshold (indicated as $t_h$) in  figure~\ref{fig:algo}C. 

As shown in figure~\ref{fig:algo}C, there are two cases where the updated data point can lie; one is that the "execution time" attribute is decreased which is indicated as $D_1$. In the second case, there is an increase in the "execution time" beyond the time threshold $T_h$ as indicated by $D_2$ that is approximated as a slowdown. Therefore, a decision to initiate performance testing is made by the algorithm. This operation is shown on line 3-5 of algorithm~\ref{algo:decision}.

On the other hand, if the "number of executed statement" attribute of the updated data point does not match with the "number of executed statement" attribute of any previous data point then time threshold is identified based on the location of  two data points which are above and below the updated data point. Moreover, data points (above and below) that have maximum ``execution time'' attribute are used to determine the time threshold; this is done by calculating the mid point of the ``execution time'' attribute of the data points above and below  shown as $T_{1_h}$ for $D_1$ or $T_{2_h}$ for $D_2$ in figure~\ref{fig:algo}D. After setting the time threshold, same conditions are applied to make a decision i.e., does the execution time attribute of updated data point cross the time threshold value as indicated for $D_2$ in figure~\ref{fig:algo}D. This operation is accomplished in line 7,12-15 of algorithm 2.

%based on the observation that the number of executed statements have a correlation with execution time (we demonstrate this in evaluation section). This means that as the number of executed statements increase, the execution time also increases and vice versa. 

%\textit{how a time threshold can be set?}. Here, identification of time threshold is based on the observation that the number of executed statements have a correlation with execution time (we demonstrate this in evaluation section). This means that as the number of executed statements increase, the execution time also increases and vice versa. 

%Based on this observation, the location of  two data points are identified in the current cluster that are above and below the updated data points. Moreover, data points (above and below) that have maximum execution time attribute are used to determine the time threshold; this is done by calculating the mid point of execution time attribute of the data points above and below  shown as $T_1_h$ for $D_1$ or $T_2_h$ for $D_2$ in figure~\ref{fig:algo}D. After setting the time threshold, same conditions are applied to make a decision i.e., does execution time attribute of updated data point crosses the time threshold value as indicated for $D_2$ in figure~\ref{fig:algo}D. This operation is accomplished in line 7,12-15 of algorithm 2.

\begin{figure*}[ht!]

    \centering
	\includegraphics[scale=0.43
	]{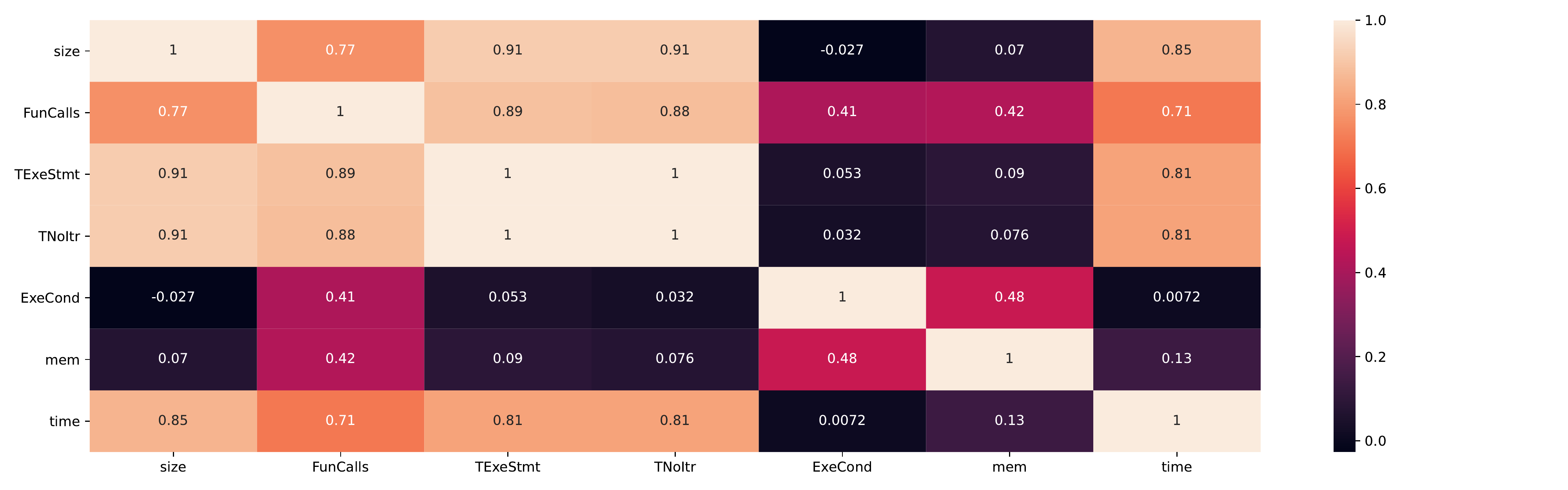}
	\caption{Correlation matrix between different variables; where labels such as \textbf{Size} represents Input size, \textbf{FunCalls} represents total function calls, \textbf{TExeStmt} represents total executed statements, \textbf{TNoItr} represents number of iterations, \textbf{ExeCond} represents executed conditionals, \textbf{Mem} represents memory used during execution of test suite, and \textbf{Time} represents the total execution time}
	\label{fig:correlation}
\end{figure*}

Furthermore, figure~\ref{fig:algo}D demonstrates a case where ``execution time'' attribute of the updated data point is close to the time threshold. Even though there is an increase in the execution time for the data point but algorithm will not decide in favor of initiation performance testing from this observation because the updated data point does not cross the time threshold. Hence missing out on a performance test opportunity.

To handle this issue, the gradient of the lines from the origin to either data points (i.e., original and updated) are measured. This is shown in figure~\ref{fig:algo}E as G and $G_1$. To check whether there is a decrease of gradient from G to $G_1$, a change is measured which is indicated as $C$ in figure~\ref{fig:algo}E. This allows one to estimate the slow-down of the time between the original and updated data point. In section~\ref{sec:recommendation}, we demonstrate the advantage for using gradient in making decisions for performance testing.

Our algorithm for slow-down detection also takes outliers into account i.e., the updated data point falls outside of the boundary of cluster. Depending on where the updated data point lies, the data point in the cluster closet to the updated data point is selected. For example,  the data point having maximum execution time attribute below the outlier data point is selected as shown in figure~\ref{fig:algo}F and the operation is shown on line 8-10 in algorithm 2.

Therefore, cluster data point evaluation algorithm (i.e., algorithm 2) identifies the position of updated data point with reference to original data point. The decision making is then done by Decision to Perform Tests algorithm. To make a decision, the algorithm will always first check whether time threshold condition is met line 2-3. On a false condition, it will measure gradients of the data points and identify whether the change is within the acceptable limit or not. Acceptable limit is the limit on slow-down i.e., how much slow-down is acceptable for the software. This parameter can be tuned by the developer.

\section{Evaluation}
\label{sec:evaluation}

The analysis of this study has been conducted on a SNIC science cloud~\cite{snic} with Ubuntu Linux 16.04.4~LTS operating system. Furthermore, the configuration of the virtual machine is 4x 2.2~GHz vCPUs, 8~GB of RAM. 

We use Gitlab CI pipeline at CERN~\cite{gitlabcern} for executing test suite. The web service that we have used for our case study is CMS uploader service which is developed at CERN~\cite{cmsuploader}. Furthermore, time measurements are taken an average of 3 repeated runs.

In our evaluation, we will begin by demonstrating the relationship between the variables (i.e., attributes) used for clustering. We used correlation test which is most widely used statistical measure to assess relationships among variables. %Furthermore, we also conduct a statistical significance test (P-value) to draw conclusion about of relationship on entire population.

Secondly, we demonstrate minimization of test inputs by applying different clustering algorithms. Finally, we use two different code updates for demonstrating the feasibility of our recommendation approach. 

%Furthermore, we also demonstrate whether elbow method is feasible in identifying initial number of clusters changing data set. We also show the accuracy of clusters identified by different clustering algorithms. Finally, we demonstrate how our approach can assist analyst during analysis of minimization by existing unsupervised learning approach.

\subsection {Correlation between variables representing program behavior}
In figure~\ref{fig:correlation}, we demonstrate a correlation between different variables that are selected by profiling. These variables represents the behavior of the code during the execution of test suite. The relationship in figure~\ref{fig:correlation} is based on executing test suite with 4000 different test inputs. This is because addition of more test inputs was not changing the behavior of our analysis. This is demonstrated in figure~\ref{fig:P-test}.

We can see from the figure that 5 out of 7 variables (or attributes) demonstrates a positive correlation i.e., variables move in the same direction.  However, there is also a negative correlation between two variables; this is between input size and executed conditions. This is quite intuitive since increasing input size should not have an effect on how conditionals are executed since conditionals are based on the evaluation of expressions. Furthermore, we also see that other variables have negligible ( or very low) correlation with executed conditionals. 

Furthermore, input size demonstrates a strong correlation with function calls, executed statements, iterations as well as execution time. This means that increasing input size will increase the number of fuction calls, executed statements, number of loops and executed time of the test suite. On the other hand, there is a weak correlation between input size and memory usage.

%we observe a correlation of input size with memory but does not exhibit a strong  relationship. 

This indicates that CMS uploader service is not a memory intensive application. We try to see whether the correlation are statistically significant by conducting a P-value test. This is shown in appendix-B.

%We will now apply different clustering algorithms to demonstrate the minimization of test data. Furthermore, we will show why our proposed approach would be suitable for developers (especially at CERN) who are dealing with ever growing data.

\begin{figure}[ht!]
    \centering
	\includegraphics[scale=0.26]{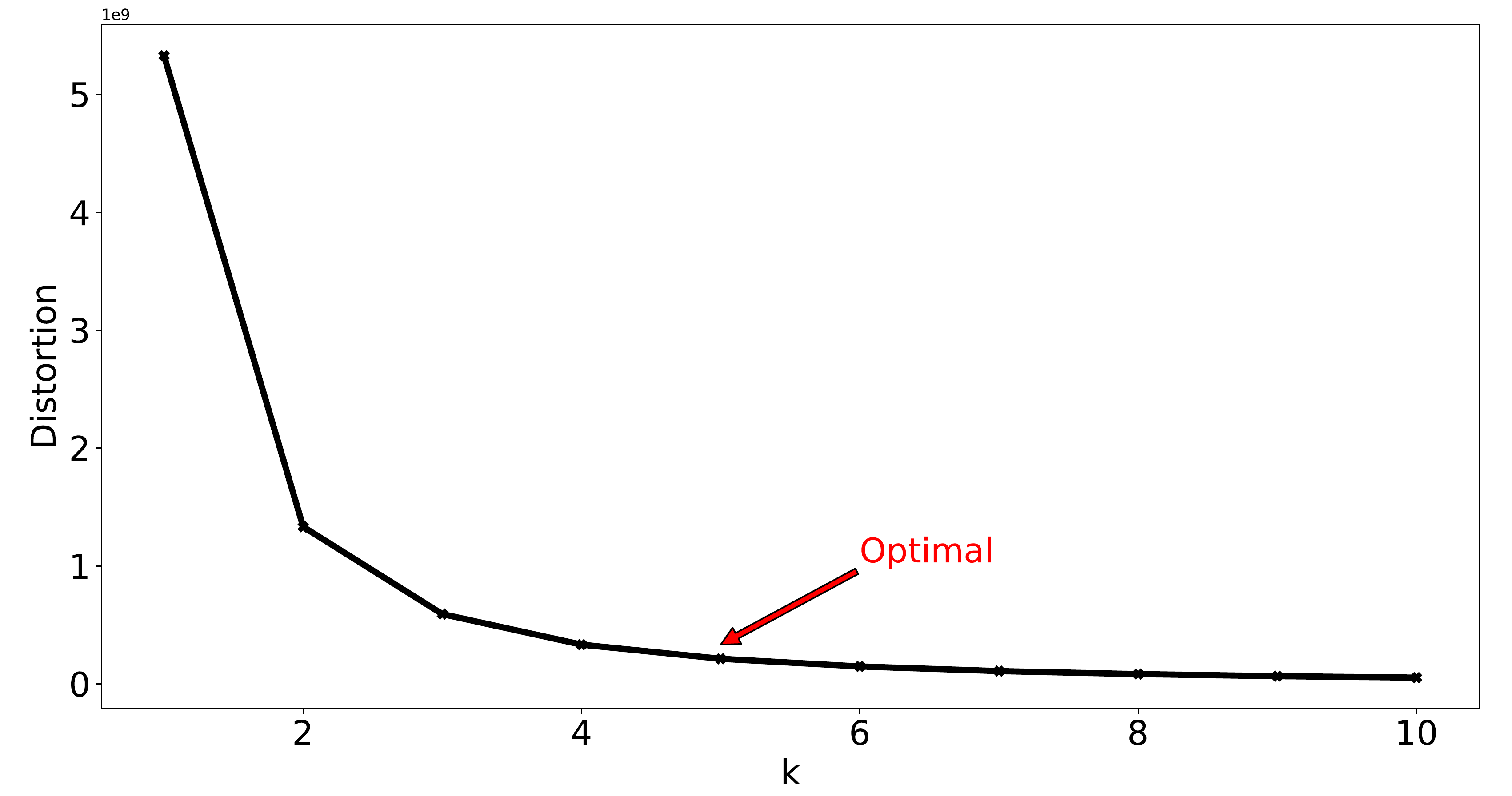}
	\caption{Elbow method for determining the number of clusters in dataset}
	\label{fig:elbow}
\end{figure}

\subsection{Identification of clusters}
We use clustering to minimize inputs based on similar code behavior. We apply most popular distance-based clustering such as K-means~\cite{kmeans}. Moreover, identification of the optimal number of clusters is done by a heuristic technique called the elbow method~\cite{elbow}. 

\begin{figure}[ht!]
%\hspace*{-1.5cm}
    \centering
	\includegraphics[scale=0.5]{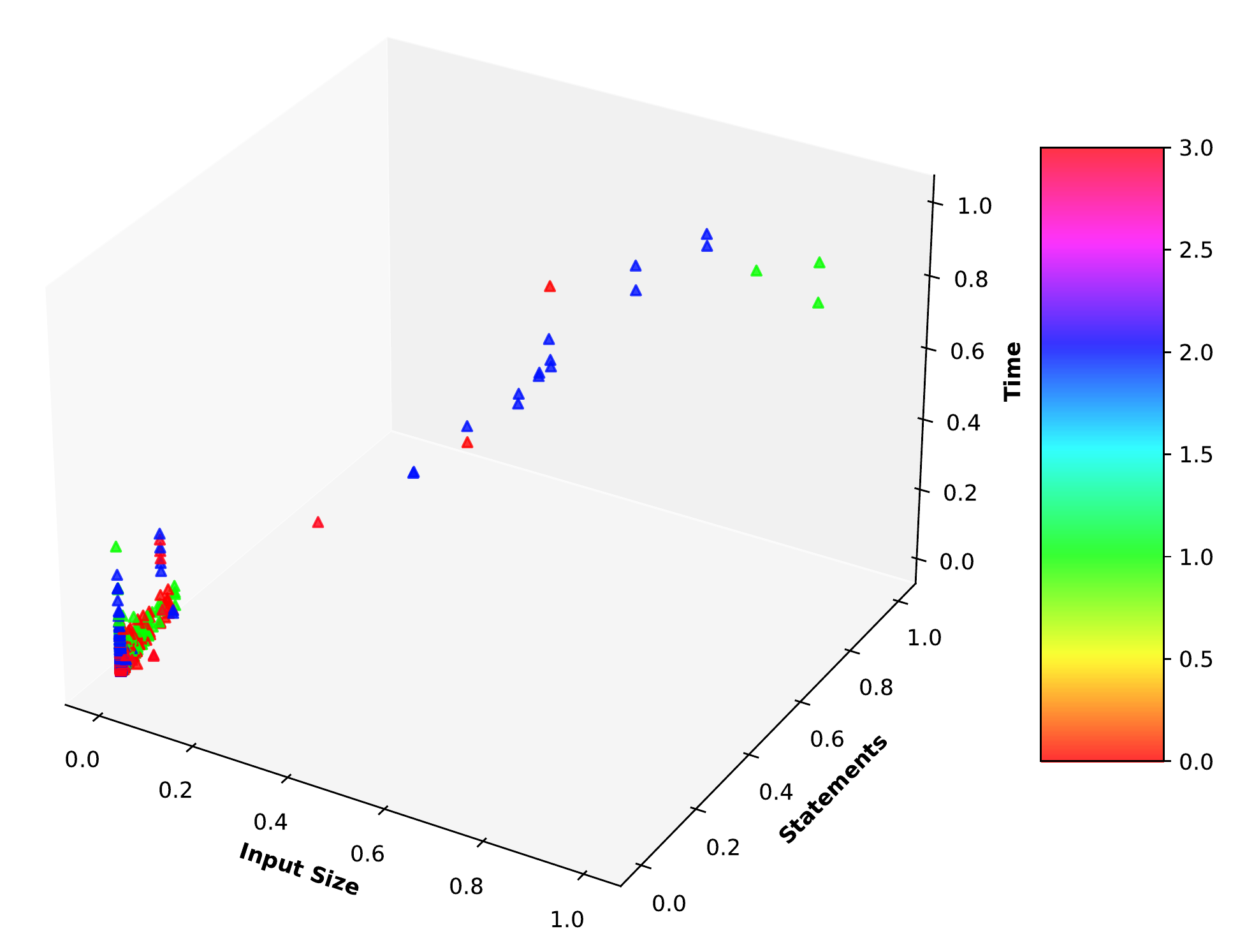}
	\caption{Three dimensional view of clustering done by distance-based clustering algorithm }
	\label{fig:cluster3d}
\end{figure}

\begin{figure}[ht!]
    \centering
	\includegraphics[scale=0.3]{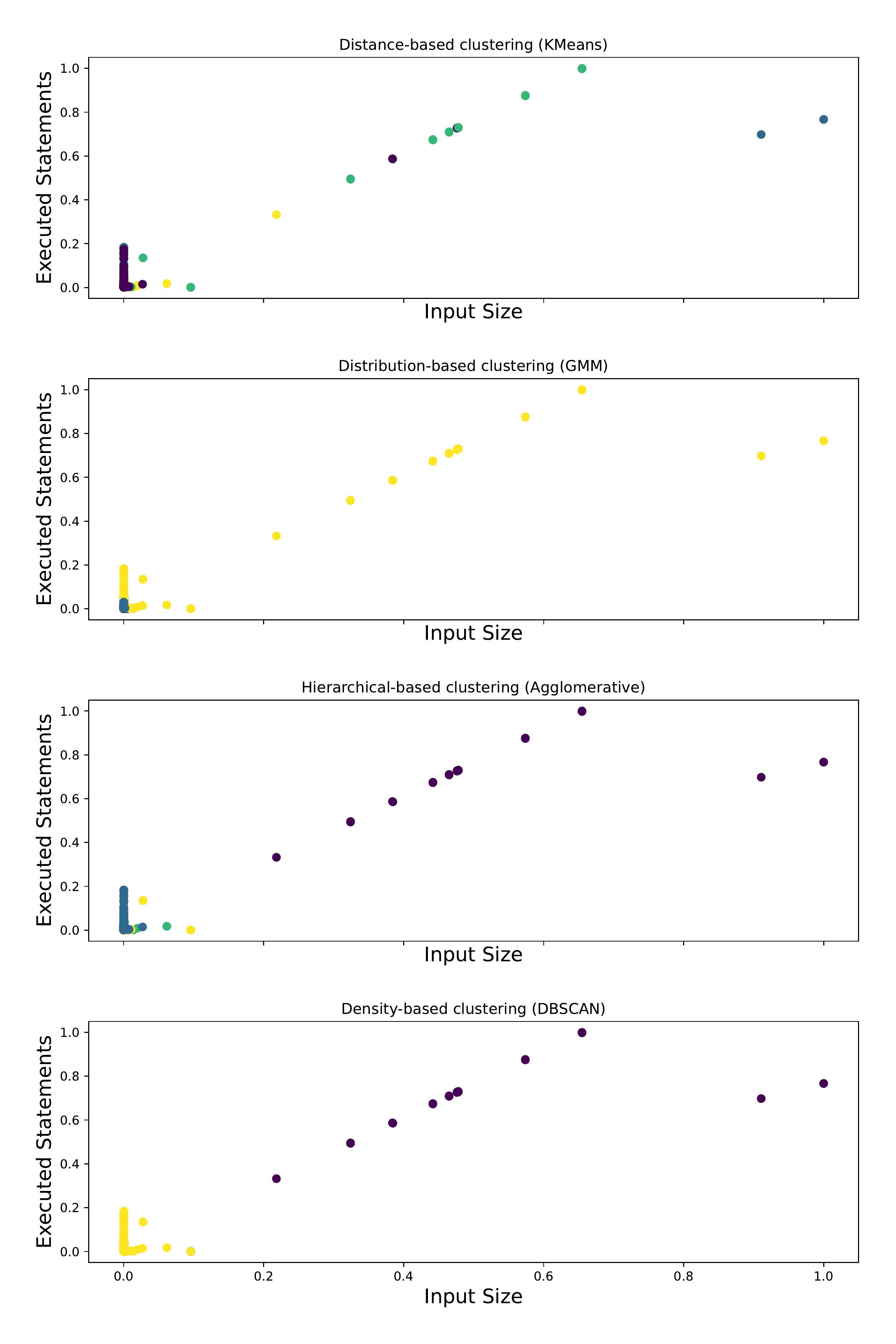}
	\caption{Different Clustering algorithms demonstrating the number of clusters}
	\label{fig:clusteralgo}
\end{figure}

This method consists of plotting the variations (or distortions) representing how well the dataset is clustered by the model. Therefore, the number of clusters is selected when the variations stop or the curve becomes smooth. Figure~\ref{fig:elbow} shows the elbow method applied on our dataset. It can be seen that distortion is minimized at 5 which is shown with an red arrow in figure~\ref{fig:elbow}.

After identifying optimal number of clusters by elbow method the result of clustering by KMeans is shown in figure~\ref{fig:cluster3d}. The figure shows a 3D scattered plot based on 3 dimensions such as input Size, executed statements and execution time. It can be seen that proper clusters are not formed based on five clusters with Kmeans. This is because clustering in KMeans is an iterative process; user has to update the number of clusters iteratively. Furthermore, the data in figure~\ref{fig:cluster3d} shows that the more appropriate number of clusters can be either 2 or 3. We will now apply different clustering algorithms and check which of the algorithm provides a better clustering.

%Commit-1 is the oldest update whereas commit-5 is the latest update to the input. We can see that the stability of curve changes in all commits e.g., the variations stops at 5 in commit-1, 4 in commit-2, 4 in commit-3, 6 in commit-4 and 5 in commit-5. Therefore, it is difficult to have a consistent identification of number of clusters. In such changing situation, this will require user's guess to select the number of clusters. Hence, we select 4 as the initial number of clusters (a median value).  Now, we will apply different clustering algorithms to see if the numbers of clusters selected as adequate for clustering.

\subsection{Clustering based on different algorithms}

Figure~\ref{fig:clusteralgo} shows clustering applied based on different types of clustering algorithm. This is shown with a scattered plot based on two features i.e., input size on x-axis and number of executed statements on y-axis. 

We have applied 4 different clustering algorithms such as density-based (K-means), distribution-based (Gaussian mixture model), hierarchical (Agglomerative), message passing based (Affinity Propagation) and Density-based (DBSCAN). We also used Affinity propagation but we realized that it was not able to create clusters; it was reporting convergence issue. We found out that this is due to a bug which is reported in the issue page of scikit-learn repository \footnote{\url{https://github.com/scikit-learn/scikit-learn/issues/17657}}.

Figure~\ref{fig:clusteralgo} gives a comparison of different clustering algorithms. It is evident that kmeans, GMM and agglomerative shows that clusters are being overlapped. The clustering done by DBSCAN algorithm seems more appropriate. It divides the clusters into two group. 

\subsubsection{Significance of cluster inspection}

Our clustering analysis gives a good indication why we need to have an inspection in our approach. An incorrect number of cluster would not only affect our approach significantly but it will also not give a useful insight into our data. 

With DBSCAN, the clustering behavior remains consistent compared to other clustering techniques. Furthermore, clustering done by DBSCAN shows that test inputs containing very low input size will have a program behavior where fewer statement are executed, fewer iteration will be made resulting in lower execution time. On the other hand, large input size will lead to high number of statements being executed resulting in higher execution time.

We will now demonstrate our decision making algorithm by 
selecting data points from each cluster and apply changes to the code.

\section{Recommendation on Code Change}
\label{sec:recommendation}
In this section we demonstrate the feasibility of our decision making part of the approach which comprises of Algorithm 1 (decision to perform tests) and Algorithm 2 (cluster data point evaluation). We have described both algorithms and their working principle in section~\ref{sec:approach}. 

We apply our algorithms on two code changes 1) A known bug in test suite which has been taken from previous study~\cite{perfci} and 2) code review based on code analysis tool --- PyLint~\cite{pylint} which is used for error checking in the code. 

This provides a reasonable evaluation of our decision making algorithm since one update already has a bug and the other update requires changes in different source files.

\subsection{Code update 1: Bottleneck in test suite}
There is a bottleneck in a constructor which occupies 99\% of the total time taken by the unit test execution. This bottleneck slows down the overall total CI execution time  by an average of 38\%. More information about the performance bug can be found in the study~\cite{perfci}.

The idea of our algorithm is to obtain minimal information to provide an informed decision about initiation of performance tests. Therefore, three data points are selected from each cluster. We also keep the acceptable limit of slow-down to 38\% since this was the overall slow-down observed due to the performance bug in test suite.

\begin{table}[ht]
\caption{Demonstration of the decision checks for Code update 1}
\centering
\begin{filecontents*}{figure/perfcidetect.csv}
checks,C11,C12,C13,C21,C22,C23
Timethreshold,True,False,False,True,True,False
Gradient,x,True,False,x,x,True
\end{filecontents*}
\pgfplotstabletypeset[
    col sep=comma,
    string type,
    columns/checks/.style={column name=\textbf{Checks}, column type={c}},
    columns/C11/.style={column name=\textbf{C11}, column type={c}},
    columns/C12/.style={column name=\textbf{C12}, column type={c}},
    columns/C13/.style={column name=\textbf{C13}, column type={c}},
    columns/C21/.style={column name=\textbf{C21}, column type={c}},
    columns/C22/.style={column name=\textbf{C22}, column type={c}},
    columns/C23/.style={column name=\textbf{C23}, column type={c}},
    every head row/.style={before row=\hline,after row=\hline},
    every last row/.style={after row=\hline},
    ]{figure/perfcidetect.csv}
    \label{table:cu1}
\end{table} 

Table~\ref{table:cu1} shows how the two decision checks provides their verdict on the slow down. In table~\ref{table:cu1}, $C_{ij}$ refers to the cluster number and the selected inputs respectively. For example, C11 corresponds to the first cluster and first input. The cross (x) indicates that Gradient check was not required since the condition of time threshold holds true. It can be seen that 50\% of updated data points (i.e., C12, C13 and C23) did not pass the conditions for time threshold. Gradient check adds a valuable asset in the decision making since it detects the rate of slow down between previous data point and its updated one. For example, C12 and C23 has a rate of slow down of $>$50\% (i.e., 66\% for C12, and 99\% for C23)  which exceeds the acceptable limit of 38\% that has been set. Hence, our algorithm gives positive decision to initiation a performance tests based on the  decision checks of 5 out of 6 data points.

In case of C12, both checks did not pass, even though there was rate of slow down of ~20\%. Developers can tune the acceptable limit based on the need of the system. For example, in some applications, it may not be feasible to have a slow-down greater than 20\%. 

\textit{The recommendation for code update 1 is correct since there is a bottleneck in the code.}

\subsection{Code update 2: Code review with PyLint}
PyLint is a static code analysis tool for used error checking. It is a popular tool for assuring code quality that gives an overall rating to the code. Furthermore, it is also used by many developers as well as  big tech companies like Google~\cite{pylintuse}. Therefore, it is a feasible tool to use for demonstrating a code update in our case study.

After analyzing the project, PyLint reported around 78\% of errors and warnings. Most of these errors can be referred to as cosmetic bugs e.g.,  errors related to the use of imports (Import checker messages~\footnote{https://vald-phoenix.github.io/pylint-errors/}) or too many boolean expressions (Design Checker Messages). Therefore, we  identified only those errors that are related to code refactoring. For example, use of enumerate instead of iterating with range and length functions. 

We identified 7 places in the project were this error occurs and we updated the code by replacing range() and len() with enumerate(). The reason for selecting this type of "error" is two-fold; firstly, we don't want to refactor the code too much such that it introduce additional errors, and secondly enumerate has an impact on the execution time.

Table~\ref{table:cu2}  shows that none of the conditions for TimeThreshold and Gradient are met i.e., all the conditions are False. Furthermore, we also observed the rate of slow-down remained under 0.5\%. 

\begin{table}[ht]
\caption{Demonstration of the decision checks for Code update 2}
\centering
\begin{filecontents*}{figure/pylintdetect.csv}
checks,C11,C12,C13,C21,C22,C23
Timethreshold,False,False,False,False,False,False
Gradient,False,False,False,False,False,False
\end{filecontents*}
\pgfplotstabletypeset[
    col sep=comma,
    string type,
    columns/checks/.style={column name=\textbf{Checks}, column type={c}},
    columns/C11/.style={column name=\textbf{C11}, column type={c}},
    columns/C12/.style={column name=\textbf{C12}, column type={c}},
    columns/C13/.style={column name=\textbf{C13}, column type={c}},
    columns/C21/.style={column name=\textbf{C21}, column type={c}},
    columns/C22/.style={column name=\textbf{C22}, column type={c}},
    columns/C23/.style={column name=\textbf{C23}, column type={c}},
    every head row/.style={before row=\hline,after row=\hline},
    every last row/.style={after row=\hline},
    ]{figure/pylintdetect.csv}
    \label{table:cu2}
\end{table}

\textit{For code update 2, the recommendation is to skip performance testing.}

\subsection{Importance of data point sampling}

Developers can be skeptical about the use of recommendation for bug identification since a miss opportunity of a performance bug detection can lead to dire consequences and their fix can be quiet time-consuming and costly~\cite{fixissue}. In this regard, developers can face a situation where  decision is 45-55\% i.e., 45\% of the data points are in favor of initiating performance test  and 55\% are in favor of skipping the tests. 

To increase developers confidence, our approach allows developers to increase the sample size of the data points during sampling stage (section~\ref{sec:sample}), and see if the decision changes i.e., more data points are in favor of initiating performance tests. However, there is a trade-off in increasing the sample size. Every time a new sample is used; a quick test is need to be run. Therefore, it is important to select sample size wisely when there is a doubt in recommendation.

\section{Conclusion}
\label{sec:conclusion}
Software performance testing is a crucial software quality assurance mechanism. However, performance testing is a time-consuming process since it requires a set of inputs to exercise different code sections. Therefore, running performance tests on every code update makes it challenging for developers because the time it takes for testing to complete. Furthermore, this becomes impractical in organizations that deals with (ever) growing data i.e., more test inputs are added to test the code.
To overcome these difficulties, we propose an approach that recommends developers when to run or skip performance tests by using minimal information from the code. We demonstrate the feasibility of our approach by applying it to two code updates and we found that our approach makes sound recommendations.

\appendices
\section{CI analytics system}

\begin{figure}[h!]
	\centering
	\includegraphics[scale=0.32]{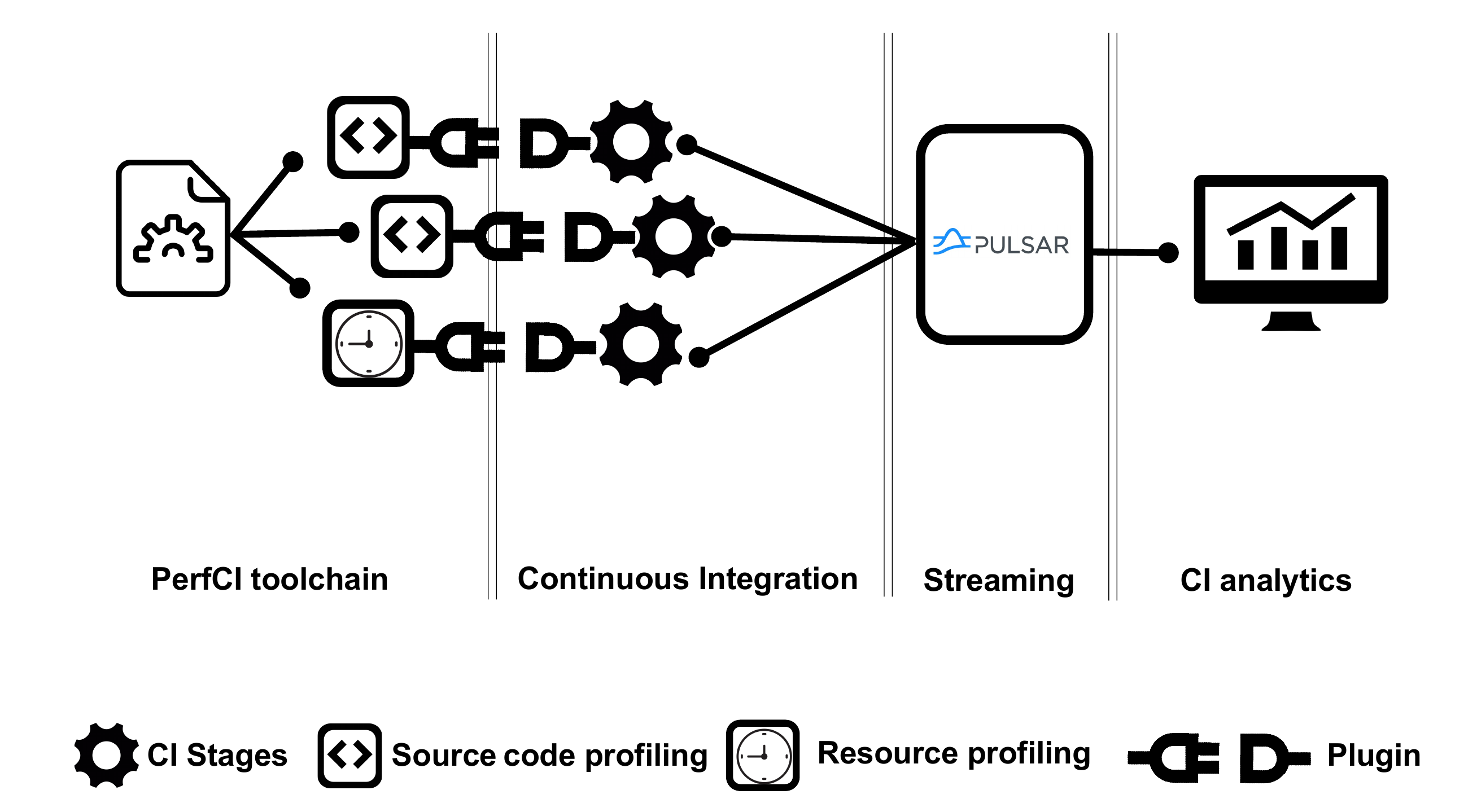}
	\caption{Overview of CI analytic system}
	\label{fig:system}
\end{figure}

Figure~\ref{fig:system} shows the overall architecture of our CI analytic system. There are four key stages of the architecture such as PerfCI toolchain, Continuous Integration, Streaming and Analytic. We briefly explain each of these four stages. 

\textit{Stage 1 (PerfCI toolchain):} We built our system on top of our previous work --- PerfCI~\cite{perfci} which allows developers to write pluggable user-defined analysis for Continuous Integration. The structure of PerfCI analysis is shown in the following code fragment. Developer implement three methods which are in the abstract class \textbf{InstrumentEvents}. 
The analysis is added to the test suite by instrumenting test cases and rewriting the CI configuration file.

\begin{lstlisting}[language=Python, frame=single]
class ResourceCollector(IntrumentEvents):
    def start_measurement(self):
        ....
    
    def end_measurement(self):
        ....
    
    def record_data(self, *args):
        ....
    
\end{lstlisting}

\textit{Stage 2 (Continuous Integration):} When the CI process runs, the $start\_measurement()$ and $end\_measurement()$ methods are executed on per unit test basis which gathers the resource information such as execution time and memory utilization. Furthermore, code-level metrics such as number of function calls are also gathered based on the PerfCI analysis defined by the user.

\textit{Stage 3 (Streaming):} One of the problems with Continuous Integration services is that it does not allow the storage of large volume of data (i.e., maximum storage size is 1GB). This is only allowed for limited time duration i.e., it will delete the storage information after a week. To overcome this issue, we use streaming framework such as Apache pulsar~\cite{pulsar} which streams our profiling data and offloads it out of the continous integration service to a remote server. This provides two advantages 1) one can perform real-time analysis and 2) large volumes of data can be handled. Furthermore, we use Apache pulsar because of its multi-tenancy feature which allows for increase scalability, cost reduction and improve security. This is quite crucial for organization like CERN.

\begin{figure}[h!]
	\centering
	\includegraphics[
	scale=0.66]{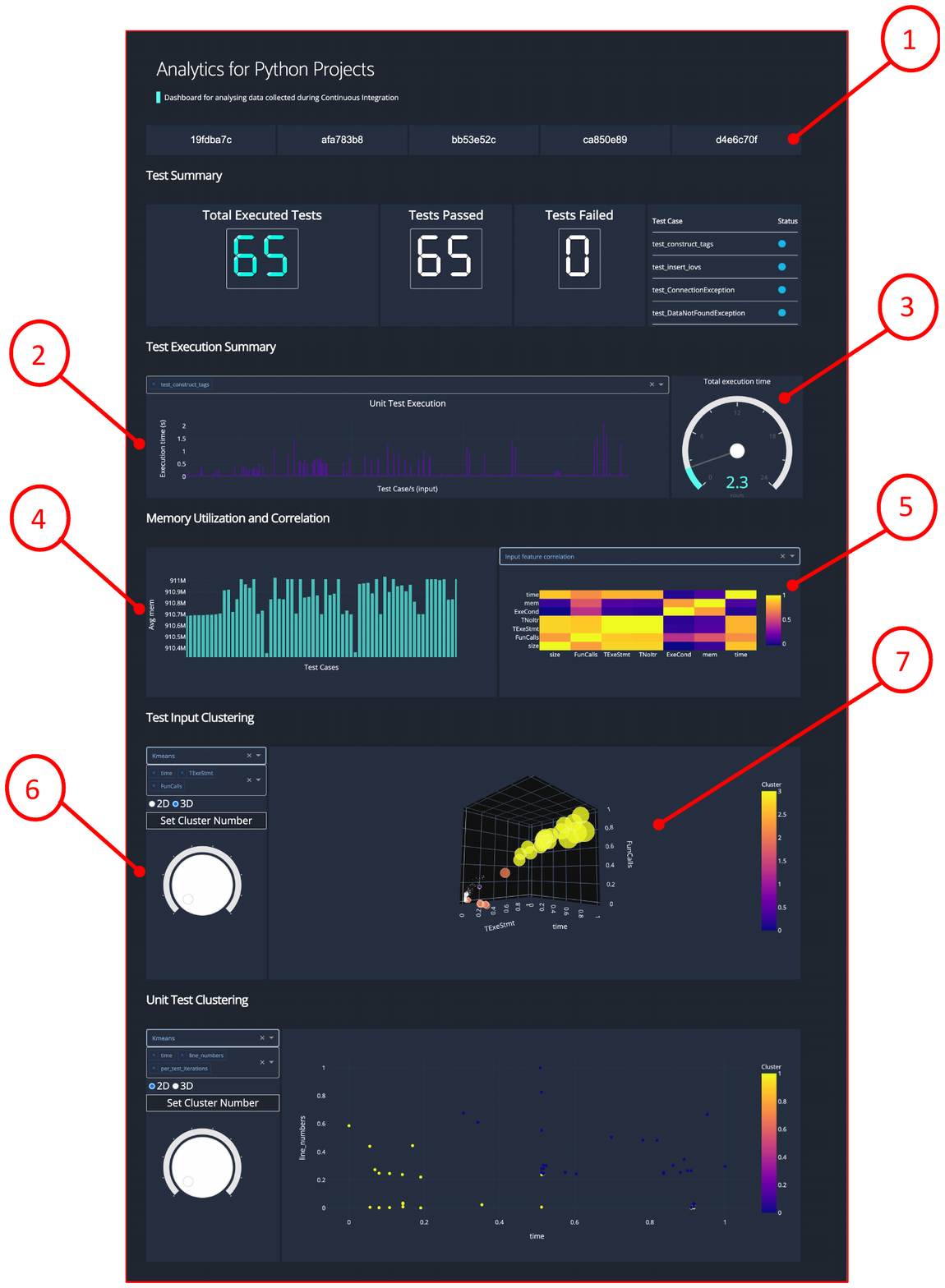}
	\caption{Analytic Dashboard}
	\label{fig:dashboard}
\end{figure}

\textit{Stage 4 (Analytics):} The analytic system has a front-end which is implemented in plotly dash~\cite{}. This allows the development of interactive visualization. The main advantage of dashboard is that the developers can visualize the clustering data i.e., inspection. Once the developer is satisfied with the clustering information. S/he can apply sampling, sample profiling and slow down detection to evaluate whether performance testing is required. This is one of the main advantage of our CI analytic system.

We now highlight some important features of our CI analytic system's dashboard. These are labeled in numbers in figure~\ref{fig:dashboard}. We briefly explain each of these numeric labels as follows:

\begin{itemize}
\item[1.] Different code updates. These are displayed as commit hash. This allows developers to observe changes in the test data on each update.

\item[2.] Execution time of each unit tests. Different unit test's execution time can be compared.

\item[3.] Total execution time of test suite. This gives a quick overview of the performance of the test at each update.

\item[4.] Memory utilized during the execution of each test case.

\item[5.] To identify correlation between different variables. 

\item[6.] Cluster Panel which provides different options such as selection of different clustering algorithm, changing visualization between 2D and 3D. This is an important feature of our dashboard that is used for inspection.

\item[7.] Three-dimensional view of clustering data. This can be used by developers if there is difficulty in identifying clear clusters.

\end{itemize}

\section{P-value test}

\begin{figure*}[ht!]
    \hspace*{-1.9cm}
    \centering
	\includegraphics[scale=0.85
	]{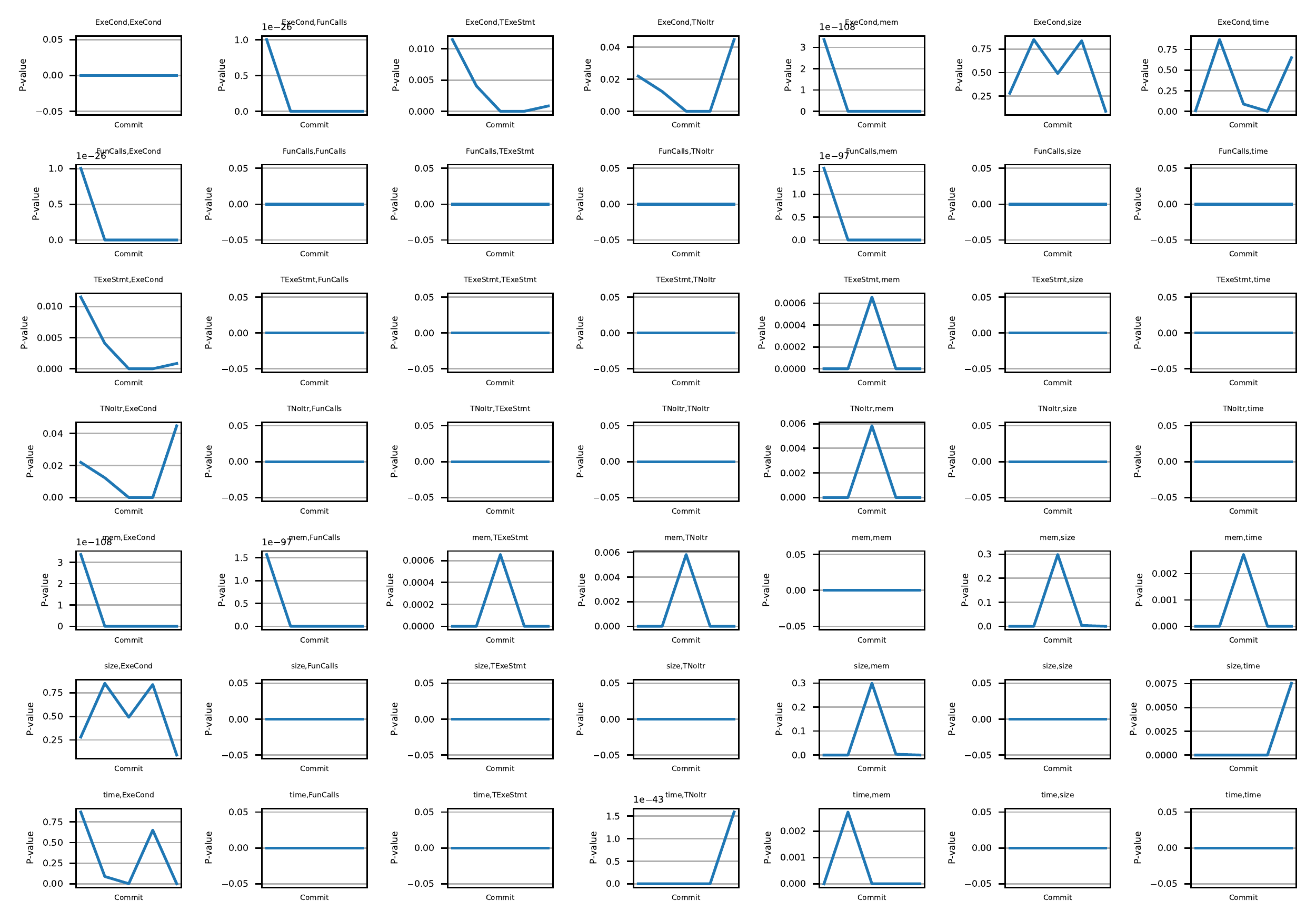}
	\caption{Statistical signficant test (P-value) for variables in evolving inputs. }
	\label{fig:P-test}
\end{figure*}

Figure~\ref{fig:P-test} shows the trend of P-values over 5 different updates to test inputs. This test is applied on the variables whose correlation is presented in section IV-A. On x-axis are the five commits i.e., updates to inputs and Y-axis shows the P-values. The title represents the two variables whose p-values are being calculated. The trends in diagonal (starting from top-left) should be ignored as they represent statistical significance between same two variables or attributes.  Variables which demonstrate high correlation also has high stable statistical significance. %P-value trend of input size and execution time changes but it is below 0.05 (indicating significance). 
Furthermore, significance of time and number of iterations change in the last commit, but the value is still well close to zero (i.e., the value is in the order of $1.5\mathrm{e}{-43}$). Therefore, there is high statistical significance for the variables representing program behavior, even with increasing the number of test inputs. This analysis provides confidence about the variable or attributes that we have used. We will now minimize the number of test inputs by grouping attributes that represents program behavior based on test input.

% or


% % use section* for acknowledgment
% \ifCLASSOPTIONcompsoc
%   % The Computer Society usually uses the plural form
%   \section*{Acknowledgments}
% \else
%   % regular IEEE prefers the singular form
% \fi


\ifCLASSOPTIONcaptionsoff
  \newpage
\fi



\bibliographystyle{IEEEtran}
\bibliography{main}

\end{document}